\documentclass[aps,prd,showpacs,amsmath,amssymb,preprintnumbers]{revtex4}

\newcommand{\vect}{\left ( \begin{array}{c}}
\newcommand{\evect}{\end{array} \right )}
\usepackage{epsfig}
\usepackage{psfrag}
\usepackage{graphicx}
\usepackage{dcolumn}
\usepackage{bm}
\def\fsl#1{\setbox0=\hbox{$#1$}                 
   \dimen0=\wd0                                 
   \setbox1=\hbox{/} \dimen1=\wd1               
   \ifdim\dimen0>\dimen1                        
      \rlap{\hbox to \dimen0{\hfil/\hfil}}      
      #1                                        
   \else                                        
      \rlap{\hbox to \dimen1{\hfil$#1$\hfil}}   
      /                                         
   \fi}                                         %

\begin{document}

\preprint{\vtop{\hbox{RU06-8-B}}}

\title{Heavy Quarkonium States with the Holographic Potential}

\author{Defu Hou}
\email[E-mail:~] {hou@iopp.ccnu.edu.cn}
\affiliation{Institute of Particle Physics, Huazhong Normal University,
Wuhan 430079, China}
\author{Hai-cang Ren}
\email[E-mail:~] {ren@mail.rockefeller.edu}
\affiliation{Physics Department, The Rockefeller University,
1230 York Avenue, New York, NY 10021-6399}
\affiliation{Institute of Particle Physics, Huazhong Normal University,
Wuhan 430079, China}

\begin{abstract}
The quarkonium states in a quark-gluon plasma is examined with the
heavy quark potential implied by the holographic principle. Both
the vanila AdS-Schwarzschild metric and the one with an infrared
cutoff are considered. The dissociation temperature is calculated
by solving the Schr\"o dinger equation of the potential model. In
the case of the AdS-Schwarzschild metric with a IR cutoff, the
dissociation temperatures for $J/\psi$ and $\Upsilon$ with the
U-ansatz of the potential are found to agree with the lattice
results within a factor of two.
\end{abstract}



\maketitle

\section{Introduction}

AdS/CFT duality opens a new avenue towards a qualitative or even
semi-quantitative understanding of the nonperturbative aspect of a
quantum field theory \cite{maldacena1} \cite{witten}. Because of
the isomorphism between the conformal group in four dimensions and
the isometry group of $AdS_5$ space, it is conjectured that the a
string theory formulated in $AdS_5\times S_5$ is dual to the $N=4$
supersymmetric Yang-Mills theory (SUSY YM) on the boundary. The
latter is believed to be conformal at quantum level and the global
$SU(4)$ symmetry of its $R$-charges corresponds to the isometry of
$S_5$. In particular, the low energy limit of the classical string
theory, the supergravity in $AdS_5\times S_5$ corresponds to the
supersymmetric Yang-Mills theory at large $N_c$ and large 't Hooft
coupling.
\begin{equation}
\lambda\equiv N_cg_{\rm YM}^2
\label{lambda}
\end{equation}
Various field theoretic correlation functions can be extracted from the metric
fluctuations of the gravity dual and the expectation value of a Wilson loop operator
is related to the minimum area the loop spans in the $AdS_5$ bulk.

Notable success has been made in the application of the AdS/CFT
duality to the physics of quark-gluon plasma (QGP) created by
RHIC, even though the underlying dynamics of QCD is very different
from that of a supersymmetric Yang-Mills theory \cite{DTSon1,
HLiu1, Yaffe}. Among them are the viscosity-entropy ratio
\cite{DTSon1}, the jet quenching parameter \cite{HLiu1} which are
closer to the observed values than the perturbative results. While
it is premature to conclude that every aspect of RHIC physics can
be explained in terms of SUSY YM, the AdS/CFT duality provides
unprecedented references since this is the only case where the
strong coupling properties of a quantum field theory can be
calculated reliably.

The heavy quarkonium dissociation is an important signal of the
formation of QGP in RHIC. In the deconfinement phase of QCD, the
range of the binding potential between a quark and an anti-quark
is limited by the screening length in a hot and dense medium,
which decreases with an increasing temperature. Beyond the
dissociate temperature, $T_d$, the range of the potential is too
short to hold a bound state and the heavy meson will melt. The
lattice simulation of the quark-antiquark potential and the
spectral density of hadronic correlators  yield consistent picture
of the quarkonium dissociation and the numerical values $T_d$. It
is the object of this paper to calculate $T_d$ using the heavy
quark potential of $N=4$ SUSY YM extracted from its gravity dual
\cite{maldacena2} \cite{SJRey} \cite{HLiu2}. Although the
potential model applies only in the non-relativistic limit which
is not the case when the 't Hooft coupling of the $N=4$ SUSY YM,
$\lambda$ becomes too strong, it can be
justified within the lower side of the range of $\lambda$ used in
the literature to compare AdS/CFT with the RHIC phenomenology,
i.e.
\begin{equation}
5.5<\lambda<6\pi.
\label{thooft}
\end{equation}
The upper edge is obtained by substituting into (\ref{lambda})
$N_c=3$ and the QCD value of $g_{\rm YM}$ at RHIC energy scale
($g_{\rm YM}^2/(4\pi)\simeq 1/2$) and the lower edge is based on a
comparison between the heavy quark potential from lattice
simulation with that from AdS/CFT \cite{gubser}.

We model the quarkonium, $J/\psi$ and $\Upsilon$ as a
non-relativistic bound state of a quark and an antiquark. The wave
function for their relative motion satisfies the Schr\"o dinger
equation
\begin{equation}
-\frac{1}{2m}\nabla^2\psi+V_{\rm eff.}(r)\psi = -E\psi
\end{equation}
where $m=M/2$ is the reduced mass with $M$ the mass of the heavy quark and $E(\ge 0)$ is the
binding energy of the bound state. Because of the
screening of QGP, the effective potential energy has a finite range and is temperature
dependent. The dissociation temperature of a particular state is the temperature at
which its $E$ is elevated to zero.

In the next section we shall calculate the dissociation temperature semi-analytically with
the heavy quark potential extracted from the vanila AdS-Schwarzschild metric. The case with
an infrared cutoff will be examined in the section III and the section IV will
conclude the paper.

\section{The Heavy Quark Potential from AdS/CFT}
The free energy of a static pair of $q\bar q$ separated by a
distance $r$ at temperature $T$ is given
\begin{equation}
e^{-\frac{1}{T}F(r,T)}=\frac{{\rm tr}<W^\dagger(L_+) W(L_-)>}{{\rm
tr}{<W^\dagger(L_+)>< W(L_-)>}} \label{define}
\end{equation}
where $L_{\pm}$ stands for the Wilson line running in Euclidean time
direction at spatial coordinates $(0,0,\pm\frac{1}{2} r)$ and is
closed by the periodicity. We have
\begin{equation}
W(L_\pm)\equiv Pe^{-i\int_0^{\frac{1}{T}}dtA_0(t,0,0,\pm\frac{1}{2}r)} \label{wilsonline}
\end{equation}
with $A_0$ the temporal component of the gauge potential subject to the
periodic boundary condition, $A_0(t+\frac{1}{T},\vec r)=A_0(t,\vec r)$. The
trace in (\ref{wilsonline}) is over the color indexes and $<...>$
denotes the thermal average. The symbol $P$ enforces the path
ordering. The corresponding internal energy reads
\begin{equation}
U(r,T) = =-T^2\frac{d}{dT}\left(\frac{F(r,T)}{T}\right)
\end{equation}
In the de-confined phase, the numerator of (\ref{define})
factorizes at large separation, i.e.
\begin{equation}
\lim_{r\to\infty}<W^\dagger(L_+)W(L_-)>=<W^\dagger(L_+)><W(L_-)>
\end{equation}
and therefore $\lim_{r\to\infty}F(r,T)=0$. The thermal average of a single Wilson line,
$<W(L)>$, is independent of the spatial coordinates.

Two ansatz of the potential model have been explored in the
literature\cite{karsch}: the $F$-ansatz which identifies $V_{\rm eff.}$ with
$F(r,T)$ and the $U$-ansatz which identifies $V_{\rm eff.}$ with
$U(r,T)$. The lattice QCD simulation reveals that the $U$ ansatz
produces a deeper potential well and thereby higher $T_d$ because the entropy
contribution is subtracted. This remains the case with holographic potential
as we shall see.

According to the holographic principle, the thermal average of a Wilson loop operator
\begin{equation}
W(C)=Pe^{-i\oint_C dx^\mu A_\mu(x)}
\end{equation}
in 4D $N=4$ SUSY YM at large $N_c$ and large 't Hooft coupling
corresponds to the minimum area $S_{\rm min.}[C]$ of the string
world sheet in the 5D AdS-Schwarzschild metric with a
Euclidean signature,
\begin{equation}
ds^2=\pi^2T^2y^2(fdt^2+d\vec x^2)+\frac{1}{y^2f}dy^2,
\label{metric}
\end{equation}
bounded by the loop $C$ at the boundary, $y\to\infty$, where
$f=1-\frac{1}{y^4}$. Specifically, we have
\begin{equation}
{\rm tr}<W(C)>=e^{-\sqrt{\lambda}S_{\rm min}[C]}.
\label{duality}
\end{equation}

For the numerator of (\ref{define}), $C$ consists of two parallel
temporal lines $(t,0,0,\pm\frac{r}{2})$ and the string world sheet
can be parameterized by $t$ and $y$ with the ansatz $x_1=x_2=0$
and $x_3$ a function of $y$. The induced world sheet metric reads
\begin{equation}
ds^2=\pi^2T^2y^2fdt^2+\Big[\pi^2T^2y^2\Big(\frac{dx_3}{dy}\Big)^2+\frac
{1}{\pi^2 T^2 y^2 f}\Big]dy^2.
\end{equation}
Minimizing the world sheet area (the Nambu-Goto action)

\begin{equation}
S[C]=(\pi T)\int_0^\beta dt\int_0^\infty dy\sqrt{1+\pi^4T^4y^4 f
\Big(\frac{dx_3}{dy}\Big)^2}
\end{equation}
generates two types of solutions\cite{maldacena2} \cite{SJRey}
\cite{HLiu2}. One corresponds to a single world-sheet with a
nontrivial profile $x_3(y)$,
\begin{equation}
x_3=\pm \pi Tq\int_{y_c}^y\frac{dy^\prime}{\sqrt{(y^{\prime
4}-1)(y^{\prime 4}-y_c^4)}}
\end{equation}
where $q$ is a constant of integration determined by the boundary
condition
\begin{equation}
r=\frac{2 q}{\pi T}
\int_{y_c}^\infty\frac{dy}{\sqrt{(y^4-1)(y^{\prime
4}-y_c^4)}} \label{boundary1}
\end{equation}
with $y_c^4=1+q^2$. The corresponding value of $\sqrt{\lambda}
S[C]$ is denoted by $I_1$. The other solution consists of two
parallel world sheets with $x_3=\pm\frac{r}{2}$ extending to the
black hole horizon and the corresponding value of $\sqrt{\lambda}
S[C]$ is denoted by $I_2$. The latter solution corresponds to two
non-interacting static quarks in the medium and is equal to the
denominator of (\ref{define})

The free energy we are interested in reads
\begin{equation}
F(r,T)=T{\rm min}(I,0)
\label{FreeEnergy}
\end{equation}
where
\begin{equation}
I\equiv I_1-I_2=\sqrt{\lambda}\int_{y_c}^\infty dy\left(\sqrt{\frac{y^4-1}{y^4-y_c^4}}-1\right)+1-y_c.
\label{nambu1}
\end{equation}
Inverting eq.(\ref {boundary1}) to express $r$ in terms of $q$ and
substituting the result to (\ref{nambu1}), it was found that the
function $I$ consists of two branches, The upper branch is always
positive and is therefore unstable. The lower branch starts from
being negative for $r<r_0$ and becomes positive for $r>r_0$. Both
branches joins at $r=r_c$ beyond which the nontrivial solution
ceases to exist. Numerically, we have $r_0\simeq \frac{0.7541}{\pi
T}$ and $r_c\simeq \frac{0.85}{\pi T}$. Introducing a
dimensionless radial coordinate,
\begin{equation}
\rho=\pi Tr,
\end{equation}
we find that
\begin{equation}
F(r,T)=-\frac{\alpha}{r}\phi(\rho)\theta(\rho_0-\rho),
\end{equation}
where $\alpha=\frac{4\pi^2}{\Gamma^4\left(\frac{1}{4}\right)}\sqrt{\lambda}
\simeq 0.2285\sqrt{\lambda}$. The screening factor $\phi(\rho)=-\rho I/(\pi\alpha)$ and
is shown in Fig.1a. We have $\phi(0)=1$ and $\phi(\rho_0)=0$ with
\begin{equation}
\rho_0 = 0.7541.
\label{quote}
\end{equation}
The small $\rho$ expansion of $\phi(\rho)$ is given by
\begin{equation}
\phi(\rho)=1-\frac{\Gamma^4\left(\frac{1}{4}\right)}{4\pi^3}\rho+\frac{3\Gamma^8\left(\frac{1}{4}\right)}
{640\pi^6}\rho^4+O(\rho^8). \label{expansion}
\end{equation}
On writing the wave function $\psi(\vec r)=u_l(\rho)Y_{lm}(\hat
r)$, the radial Schr\"odinger equation for a zero energy bound
state reads
\begin{equation}
\frac{d^2u_l}{d\rho^2}+\frac{2}{\rho}\frac{du_l}{d\rho}-\Big[\frac{l(l+1)}{\rho^2}+{\cal V}\Big]u_l=0
\label{radialeq}
\end{equation}
with ${\cal V}=MV_{\rm eff.}/(\pi^2T^2)$. We have
\begin{equation}
{\cal V}=-\frac{\eta^2}{\rho_0\rho}\phi(\rho)\theta(\rho_0-\rho)
\end{equation}
for the F ansatz and
\begin{equation}
{\cal V}=-\frac {\eta^2}{\rho_0\rho}\Big[\phi(\rho)-\rho\Big(\frac{d\phi}{d\rho}\Big)\Big]
\theta(\rho_0-\rho)
\end{equation}
for the U ansatz, where
\begin{equation}
\eta = \sqrt{\frac{\alpha\rho_0 M}{\pi T}}
\label{etadef}
\end{equation}
Note that the potential of the U-ansatz jumps to zero from below at $\rho=\rho_0$, since the derivative
of $\phi(\rho)$ is nonzero there.
For both ansatz, and the case with an infrared cutoff discussed below,
the solution to (\ref{radialeq}) is given by
\footnote{The exponential decay factor that ensures the normalizability of a bound state
wave function approaches to one in the limit of zero binding energy.}
\begin{equation}
u_l={\rm const.}\rho^{-l-1}
\end{equation}
at $\rho>\rho_0$ and by
\begin{equation}
u_l={\rm const.}\rho^l
\end{equation}
near the origin. The threshold value of $\eta$ at the dissociation
temperature, $\eta_d$, is determined by the matching condition at
$\rho=\rho_0$,
\begin{equation}
\frac{d}{d\rho}(\rho^{l+1}u_l)\mid_{\rho=\rho_0^-}=0.
\label{matching}
\end{equation}
It follows from (\ref{etadef}) that the dissociation temperature is given by
\begin{equation}
T_d=\frac{\alpha\rho_0M}{\pi\eta_d^2}=
\frac{4\pi\rho_0}{\Gamma^4\left(\frac{1}{4}\right)\eta_d^2}\sqrt{\lambda}M.
\label{melting}
\end{equation}

It is interesting to observe that the extrapolation of the first
two terms of (\ref{expansion}) vanishes at
\begin{equation}
\rho=\rho_0^\prime=\frac{4\pi^3}{\Gamma^4\left(\frac{1}{4}\right)}
\simeq 0.7178,
\label{extrap}
\end{equation}
which is very close to the exact zero point (\ref{quote}), and the
third term of (\ref{expansion}) remains small there. This suggests
that the screening factor $\phi(\rho)$ can be well approximated by a linear function
\begin{equation}
\bar\phi(\rho)\simeq 1-\frac{\rho}{\bar\rho_0}
\end{equation}
with $\bar\rho_0=\frac{1}{2}(\rho_0+\rho_0^\prime)\simeq 0.7359$, as is evident from
the exact profile of $\phi(\rho)$ shown in Fig.1a.  The effective potential
$V_{\rm eff.}$ is then approximated by a truncated Coulomb potential. We have
\begin{equation}
{\cal
V}=-\frac{\alpha}{\rho_0\rho}\Big(1-\frac{\rho}{\rho_0}\Big)\theta(\rho_0-\rho)
\end{equation}
for the F-ansatz and
\begin{equation}
{\cal V}=-\frac{\alpha}{\rho_0\rho}\theta(\rho_0-\rho)
\end{equation}
for the U-ansatz, where the over bar of $\rho_0$ has been suppressed.

\begin{figure}
\includegraphics[scale=0.48,clip=true]{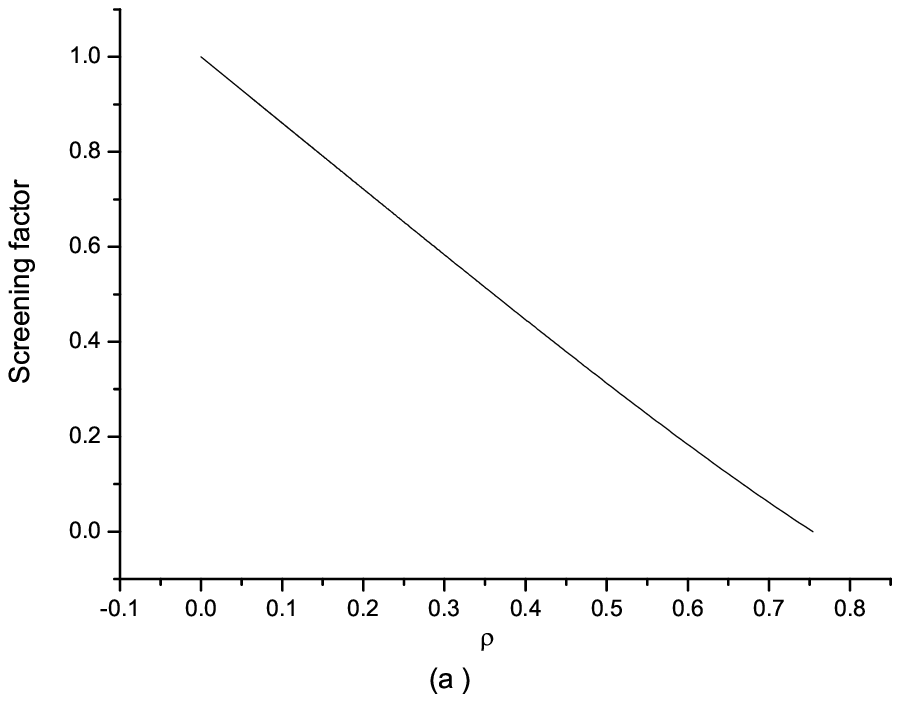}
\includegraphics[scale=0.48,clip=true]{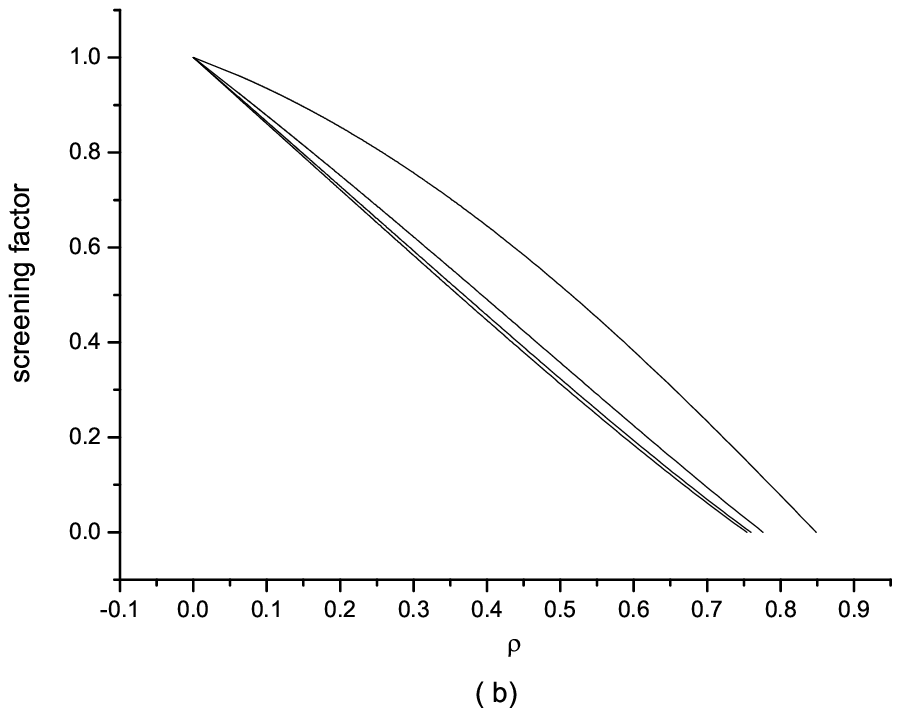}
\caption{\label{fig:epsart} (a)The exact screening factor $\phi(\rho)$ profile
extracted from the metric (\ref{metric}). (b) The screening factors
$\chi(\rho,T)$'s extracted from
the metric (\ref{metric_yee}) at different ratios of
$T/T_c=1,2,3$,$\infty$ from top to bottom. }
\end{figure}

The radial wave function of the F-ansatz under the truncated Coulomb approximation
can be expressed in terms of the confluent hypergeometric function of the 1st
kind for $\rho<\rho_0$, i.e.
\begin{equation}
u_l=\rho^l{}_1F_1(l+1-\frac{\eta}{2};2l+2;2\eta\frac{\rho}{\rho_0}).
\end{equation}
The matching condition (\ref{matching}) yields the secular equation for $\eta$,
\begin{equation}
2l+1-\eta+\eta\left(1-\frac{\eta}{2l+2}\right)
\frac{{}_1F_1(l+2-\frac{\eta}{2};2l+3;2\eta)}{{}_1F_1(l+1-\frac{\eta}{2};2l+2;2\eta)}
=0.
\end{equation}
As $\eta$ is reduced from above, we expect the bound state to melt
successively. Therefore the first positive root corresponds to the
minimum binding strength for a bound state of angular momentum $l$
and the 2nd one to the threshold of the first radial excitation of
the same partial wave. Knowing the values of these $\eta$'s, the
disassociation temperature can be computed from the formula
(\ref{melting}). For example, the threshold $\eta$ of the 1$s$
state, $\eta_{1s}\simeq 1.76$, which implies that
\begin{equation}
T_d\simeq 0.0174\sqrt{\lambda}M.
\label{scaleF}
\end{equation}

In case of the $U$-ansatz under the same approximation, we find that
\begin{equation}
u_l=\frac{1}{\sqrt{\rho}}J_{2l+1}\Big(2\eta\sqrt{\frac{\rho}{\rho_0}}\Big)
\end{equation}
for $\rho<\rho_0$ with $J_\nu(x)$ the Bessel function. The secular equation for $\eta$ reads
\begin{equation}
2l+1-\eta\frac{J_{2l+2}(2\eta)}{J_{2l+1}(2\eta)}=0.
\end{equation}
We have $\eta_{1s}=1.21$ and
\begin{equation}
T_d\simeq 0.0368\sqrt{\lambda}M.
\label{scaleU}
\end{equation}

Numerical results for the dissociation temperature of quarkonium
are tabulated in table I, where we have used the mass values $M=1.65$GeV,
$4.85$ GeV for $c$ and $b$ quarks \cite{dataB}
The  errors caused by the truncated Coulomb approximation are
within 4 percent, as is shown by the numerical solution to the
secular equation of the exact potential.

\begin{table}
\begin{tabular}{c|r|r|r|r|r|r}
ansatz\kern8pt& $J/\psi(1s)$ \kern8pt& $J/\psi(2s)$ \kern8pt& $J/\psi(2p)$
\kern8pt& $\Upsilon(1s)$ \kern8pt& $\Upsilon(2s)$ \kern8pt& $\Upsilon(2p)$ \kern8pt\\
\hline
$F$ \kern8pt&67-124 \kern8pt& 15-28 \kern8pt& 13-25
\kern8pt& 197-364 \kern8pt& 44-81 \kern8pt& 40-73 \kern8pt\\
\hline
$U$ \kern8pt& 143-265 \kern8pt& 27-50 \kern8pt& 31-58
\kern8pt& 421-779 \kern8pt& 80-148 \kern8pt& 92-171 \kern8pt\\
\hline
\end{tabular}
\bigskip
\caption{$T_d$ in MeV's under the truncated Coulomb approximation. The lower value
of each entry corresponds to $\lambda=5.5$ and the upper one to $\lambda=6\pi$.}
\label{table-IIIa}
\end{table}

To assess the validity of the potential model employed, we consider a classical
two body problem with the truncated Coulomb potential (\ref{FreeEnergy}).
The Lagrangian of the system reads
\begin{equation}
L=\frac{1}{2}M({\dot\vec r}_1^2+{\dot\vec r}_2^2)-V_{\rm eff.}(|\vec
r_1-\vec r_2|)
\end{equation}
Introducing the center of mass coordinates and the relative
coordinates via $\vec r_1=\vec R+\vec r/2$ and $\vec r_2=\vec R-\vec r/2$,
and assuming a static pair, $\dot{\vec R}=0$, we have
\begin{equation}
L=\frac{1}{4}M{\dot{\vec r}}^2-V_{\rm eff.}(r)
\end{equation}
which becomes a particle of the reduced mass in an central potential $V_{\rm eff.}(r)$.
For a circular orbit, the force balance at the border of the potential well,
$r=d\equiv\rho_0/(\pi T)$ is given by
\begin{equation}
\frac{Mv_r^2}{2d}=\frac{\alpha}{d^2}.
\end{equation}
which yields
\begin{equation}
v^2=\frac{v_r^2}{4}=\frac{\alpha^2}{2\eta^2}
\end{equation}
with $v^2=\dot\vec r_1^2=\dot\vec r_2^2$ and $v_r=\dot\vec r$.
At melting temperature of U-ansatz, we find that $0.119<v^2<0.408$ for $\lambda$ in
the range of (\ref{thooft}). Therefore the
approximation may be problematic at the upper limit of the range
(\ref{thooft}). On the other hand, the lower values of $\alpha$
was advocated in \cite{gubser} based on a comparison between the
potentials from lattice simulation and that from the AdS/CFT and
may serve our purpose better. The NR approximation works better
for the dissociation temperature of excitations because of the
higher $\eta$ values.

\section{Quarkonium in a Model with an Infrared Cutoff}

Because of the conformal invariance at quantum level, there is no
color confinement in $N=4$ SUSY YM even at zero temperature. In order to
simulate the confined phase of QCD at low temperature, an infrared cutoff
has to be introduced that suppress the contribution of the $AdS$ horizon.
Two scenarios explored in the literature are the hard-wall model and the soft-wall models.
The gravity dual of de-confinement transition is modeled as the Hawking-Page
transition from a metric without a black hole to that with a black
hole with $T>T_c$. The gravity dual of the free energy with a hard wall is the
Einstein-Hilbert action with a cosmological constant given by

\begin{equation}
F=-\frac{T}{16\pi G_5}\int d^4 x\int_0^{z_0} dz\sqrt{g}(R-12),
\label{hard}
\end{equation}
subject to an appropriate UV regularization, where $R$ is the
curvature scalar. In the hadronic phase, the metric underlying $g$
and $R$ is that of the standard $AdS_5$
\begin{equation}
ds^2=\frac{1}{z^2}(dt^2+d\vec x^2+dz^2),
\label{metric_ads}
\end{equation}
truncated beyond $z_0$ with $z_0$ determined by the $\rho$-meson
mass \cite{DTSon2}. In the plasma phase, the underlying metric is
given by
\begin{equation}
ds^2=\frac{1}{z^2}(fdt^2+d\vec x^2+f^{-1}dz^2),
\label{metric_adsbh}
\end{equation}
with $f=1-\pi^4T^4z^4$, which is identical to (\ref{metric}) upon
the transformation $y=1/(\pi T z)$. The domain of the
$z$-integration is $0<z<1/(\pi T)$ which corresponds to
$1<y<\infty$. Notice that $z_0>1/(\pi T)$ above the transition
temperature $T_c$, which was found to be $T_c\simeq 0.1574m_\rho$
\cite{herzog}. Therefore heavy quark potential and the meson
dissociation temperatures calculated above are identical to what
calculated above for the vanila $N=4$ SUSY YM.

In case of the simplest soft-wall model (\cite{DTSon3}), a dilaton is introduced with
eq.(\ref{hard}) modified to
\begin{equation}
F=-T\frac{1}{16\pi G_5}\int d^4x\int_{\rho_0}^\infty dr
e^{-\frac{c}{\rho^2}}\sqrt{g}(R-12), \label{soft1}
\end{equation}
with $c$ determined by the $\rho$-mass and the transition
temperature is predicted as $T_c\simeq 0.2459m_\rho$
\cite{herzog}. The string frame metric underlying $g$ and $R$
remains given by (\ref{metric_ads}) in the hadronic phase and by
(\ref{metric_adsbh}) in the plasma phase. Although, the infrared
cutoff is partially carried over to the plasma phase, the
heavy-quark potential and the dissociation temperatures thus
obtained remains intact since the minimum area dual to a Wilson
loop has to be defined with respect to the string frame metric
\footnote{ We are obliged to James T. Liu for pointing it out to
us}.

A variant of the soft-wall scenario proposed in ref.\cite{JTYee}, however, admits a
string frame metric that is different from (\ref{metric_adsbh}) by a conformal factor, i. e.
\begin{equation}
ds^2=\frac{e^{bz^2}}{z^2}(fdt^2+d\vec x^2+f^{-1}dz^2),
\label{metric_yee}
\end{equation}
The value of $b=0.184{\rm GeV}^2$ was obtained
by fitting the lattice simulated transition temperature $T_c=186$MeV. Following the steps
from (\ref{metric}) to (\ref{nambu1}), we find that
\begin{equation}
F(r,T)=-\frac{\alpha}{r}\chi(\rho, T)
\label{holographic}
\end{equation}
where the screening factor $\chi(\rho,T)$ is defined
parametrically by
\begin{equation}
\chi=-\frac{\rho\sqrt{\lambda}}{\pi\alpha}\Big[\int_{y_c}^\infty dy e^{
\frac{\beta}{y^2}}
\left(\sqrt{\frac{y^4-1}{y^4-1-q^2e^{-\frac{2\beta}{y^2}}}}-1\right)-
\int_1^{y_c}dye^{\frac{\beta}{y^2}}\Big]. \label{nambu}
\end{equation}
and
\begin{equation}
\rho=\frac{2q}{2T} \int_{y_c}^\infty \frac{dy e^{-\frac{\beta}{y^2}}}
{\sqrt{(y^{ 4}-1)(y^{ 4}-1-q^2e^{-\frac{2\beta}{y^2}})}}
\label{boundary}
\end{equation}
with
\begin{equation}
q^2 = (y_c^4-1)e^{\frac {2\beta}{y_c^2}}
\end{equation}
and $\beta=\frac{b}{\pi^2T^2}=0.539\frac{T_c^2}{T^2}$.
We have $\chi(\rho,\infty)=\phi(\rho)$.
The small $\rho$ behavior of $\chi(\rho,T)$ reads

\begin{equation}
\chi(\rho,T)=1-\frac{\Gamma^4\left(\frac{1}{4}\right)}{4\pi^3}a\rho+O(\rho^2).
\label{soft}
\end{equation}
where
\begin{equation}
a=e^{\beta}-2\sqrt{\beta}\int_0^{\sqrt{\beta}}dx e^{x^2}.
\label{afactor}
\end{equation}
The numerical results of $\chi(\rho,T)$ for several values of the
ratio $T/T_c$ are shown in Fig. 1b.
As is seen, the screening becomes weaker in the neighborhood of
$T_c$, similar to lattice QCD \cite{karsch}. The function $\chi(\rho,T)$
vanishes at $\rho_0\simeq
0.8485$ at $T=T_c$ while the extrapolation of the first two terms
of (\ref{soft}) vanishes at
$\rho_0^\prime\simeq 1.764$. The truncated Coulomb approximation
deteriorates in the vicinity of $T_c$.

To determine the dissociation temperature in this case, we solve
the Schr\"o dinger equation (\ref{radialeq}) numerically with the
numerically calculated heavy quark potential for both ansatz. We
have
\begin{equation}
{\cal V}=-\frac{\eta^2}{\rho_0\rho}\chi(\rho,T)
\end{equation}
for the F ansatz and
\begin{equation}
{\cal V}=-\frac {\eta^2}{\rho_0\rho}\Big[\chi(\rho,T)-\rho\Big(\frac{\partial
\chi}{\partial\rho}\Big)_T -T\Big(\frac{\partial\chi}{\partial
T}\Big)_\rho\Big]
\end{equation}
for the U ansatz. The solution for $\rho<\rho_0$ can be obtained
by standard Runge-Kutta method and the threshold value of $\eta$,
$\eta_d$ follows from the matching condition (\ref{matching}).
Notice that the the infrared cutoff renders both $\rho_0$ and
$\eta_d$ nontrivial functions of temperature and
eq.(\ref{melting}) becomes an {\it implicit} equation of $T_d$. Nor does
$T_d$ scales simply with $\sqrt{\lambda}$ and $M$ according to (\ref{scaleF})
and (\ref{scaleU}.
The modified dissociation temperatures are tabulated in the table
II, which show an significant increment in the dissociation temperature in the
vicinity of $T_c$. The comparison between the ratios $T_d/T_c$ we calculated
here with that obtained from the lattice QCD is shown in table III.

\begin{table}
\begin{tabular}{c|r|r|r}
ansatz\kern8pt& $J/\psi$ \kern8pt& $\Upsilon$ \kern8pt\\
\hline
$F$ \kern8pt& NA \kern8pt& 235-385 \kern8pt\\
\hline
$U$ \kern8pt& 219-322 \kern8pt& 459-779 \kern8pt\\
\hline
\end{tabular}
\bigskip
\caption{$T_d$ in MeV's for the 1$s$ state with the deformed
metric. "NA" means that there is no bound states above $T_c$ and
the entry for the $\Upsilon$ with $U$ ansatz and $\alpha=6\pi$ is
taken from the table I, since no significant increment is
observed.}
 \label{table-II}
\end{table}

\begin{table}
\begin{tabular}{c|r|r|r|r|r}
ansatz\kern8pt& $J/\psi$(holographic) \kern8pt& $J/\psi$(lattice)
\kern8pt& $\Upsilon$(holographic) \kern8pt& $\Upsilon$(lattice) \kern8pt\\
\hline
$F$ \kern8pt& NA \kern8pt& 1.1 \kern8pt& 1.3-2.1 \kern8pt& 2.3 \kern8pt\\
\hline
$U$ \kern8pt& 1.2-1.7 \kern8pt& 2.0 \kern8pt& 2.5-4.2 \kern8pt& 4.5 \kern8pt\\
\hline
\end{tabular}
\bigskip
\caption{The ratio $T_d/T_c$ from the holographic potential (\ref{holographic}) and that from the
lattice QCD \cite{karsch}.}
\label{table-III}
\end{table}

We observe that the longer screening length near $T_c$ requires
that $b>0$ in the string frame metric (\ref{metric_yee}). A
negative $b$ makes $a$ of (\ref{afactor}) greater than one and
$\rho_0^\prime$ smaller than the corresponding value of the
conformal (\ref{metric}). The numerical result of $\rho_0$ for
$b<0$ is also found to be smaller and the temperature dependence
of the screening length is opposite to that of the lattice QCD.

\section{Concluding Remarks}
\label{sec-summary} In summary, we have calculated the
dissociation temperatures of heavy quarkonium using the $AdS/CFT$
implied potential with both the vanila AdS-Schwarzschild metric
and the one with the infrared cutoff. While estimations of $T_d$
have been made in the literature based on various holographic
models \cite{JunLi} \cite{Kim}, a determination of $T_d$ from the
Schr\"odinger equation within the same framework is still lacking.
Our work is to fill this gap. On comparing our results with that
from the lattice simulation \cite{karsch}, we found that our
ratios $T_d/T_c$ extracted from the modified AdS-Schwarzschild
metric (\ref{metric_yee}) are lower than the lattice ones within a
factor of two. The difference between the F-ansatz and the
U-ansatz is similar. One has to bear in mind that the matter field
proportion of $N=4$ SUSY YM is larger than that of QCD with light
quarks and the screening ought to be stronger. The stronger
screening will make the heavy quarkonium more vulnerable and
thereby lower the dissociation temperature. This is consistent
with the observation that the potential becomes wider in the
metric with the IR cutoff introduced in \cite{JTYee} since some of
degrees of freedom becomes massive.

The authors of \cite{hoyos} calculated the spectral function of
the fluctuation of a D7 brane and deduced from which the meson
melting temperature
\begin{equation}
T_d=\frac{2.17M}{\sqrt{\lambda}}.
\label{spectral}
\end{equation}
for all bound state levels. On substituting the value of the heavy
quark masses, it gives rise to $825{\rm MeV}<T_d<1.53{\rm GeV}$
for $J\slash \psi$ and $2.42{\rm GeV}<T_d<4.85{\rm GeV}$ for
$\Upsilon$ within the range (\ref{thooft}) of the 't Hooft
coupling. The spectral analysis of \cite{hoyos} would be superior
if the underlying dynamics of QGP were $N=4$ SUSY YM. But the
higher $T_d$'s may point to its difference from QCD. Also the
$\lambda$ dependence of $T_d$ in eq.(\ref{spectral}) is entirely
at variance with ours. Our relationship $T_d\propto\sqrt{\lambda}$
for the vanila AdS-Schwarzschild metric follows from the property
that the binding strength is proportional to $\sqrt{\lambda}$ but
the binding range is independent of $\lambda$. A field theoretic
speculation on this property of the heavy quark potential can be
found in \cite{shuryak}. Since we are comparing two different
theories, some features may be shared by both and some properties
may not. If the screening properties of the $N=4$ SUSY YM can be
carried over to QCD, our potential model calculation inspired by
the holographic principle should provide a semi-quantitative
description on the quarkonium dissociation mechanism in the
realistic quark-gluon plasma.


\section*{Acknowledgments}
We would like to thank Mei Huang, James T. Liu, R. Mawhinney, P.
Shock and Pengfei Zhuang for discussions. The work of D.F.H. is
supported in part by Educational Committee under grants
NCET-05-0675 and project No. IRT0624.. The work of D.F.H and
H.C.R. is also supported in part by NSFC under grant No. 10575043
and by US Department of Energy under grant
DE-FG02-01ER40651-TaskB.

\end{document}